**Surface Chemistry and Electrical Properties of Germanium Nanowires**

Dunwei Wang[†], Ying-Lan Chang[‡], Qian Wang[†], Jien Cao[†], Damon Farmer[¶], Roy Gordon[¶] and Hongjie Dai[†*]

[†]Department of Chemistry, Stanford University, Stanford, CA 94305

[‡]Agilent Laboratories, Agilent Technologies, Inc. 3500 Deer Creek Road, Palo Alto, CA 94304.

[¶] Department of Chemistry and Chemical Biology, Harvard University, Cambridge, MA 02138

* hdai@stanford.edu

**Abstract:** Germanium nanowires (GeNWs) with p- and n-dopants were synthesized by chemical vapor deposition (CVD) and used to construct complementary field effect transistors (FETs). Electrical transport and x-ray photoelectron spectroscopy (XPS) data are correlated to glean the effects of Ge surface chemistry to the electrical characteristics of GeNWs.  Large hysteresis due to water molecules strongly bound to $GeO_2$ on GeNWs is revealed. Different oxidation behavior and hysteresis characteristics and opposite band bending due to Fermi level pinning by interface states between Ge and surface oxides are observed for p- and n-type GeNWs.  Vacuum annealing above 400 °C is used to remove surface oxides and eliminate hysteresis in GeNW FETs. High-κ dielectric $HfO_2$ films grown on clean GeNW surfaces by atomic layer deposition (ALD) using an alkylamide precursor is effective serving as the first layer of surface passivation. Lastly, the depletion length along the radial direction of nanowires is evaluated. The result suggests that surface effects could be dominant over the 'bulk' properties of small diameter wires.

**Introduction**

Currently, as electronic devices are scaled down to the sub-100 nm regime, Ge has gained renewed interest as a material of choice for future electronics due to its significantly higher electron and hole mobilities than Si.[1,2] It is well known that the water solubility of $GeO_2$ and the lack of a stable oxide have prevented Ge from being utilized as an electronic material in the past. This problem could now be solved, in principle, by using the recently advanced high dielectric-constant (high-κ) films as gate insulators. Indeed, promising results have already been reported for metal-oxide-semiconductor capacitors and field effect transistors (FET) based on Ge with high-κ films as dielectric materials.[3,4] Despite these progresses, the surface chemistry of Ge remains much less explored and understood than that of Si. Detailed understanding of how various species on Ge surfaces affecting the electrical properties of Ge is lacking. Further, approaches to reliably passivate Ge surfaces still have to be fully established and require significant effort.

Chemically synthesized nanomaterials such as nanotubes and nanowires have attracted much attention lately as building blocks for future electronics.[5-13] These bottom-up chemically derived materials may present advantages over top-down lithographically patterned materials in better intrinsic properties in the case of carbon nanotubes,[6,7] higher degrees of structural or surface perfection, lower costs than single crystalline wafer materials in the case of Ge or high flexibility in the type of substrates that can be used for assembly of these materials. Ge nanowires are particularly appealing due to the high carrier mobility and the ease of synthesis. We have shown recently that single crystalline GeNWs can be readily synthesized by a simple chemical vapor deposition (CVD) method at a low temperature of 275°C from Au nanoparticle seeds.[10] The growth condition for GeNWs is the mildest among all semiconducting wires and represents the lowest temperature under which single crystalline semiconductors are synthesized. The potential of the GeNWs as building blocks for electronic devices has also been demonstrated by successful fabrication of FETs based on p-type GeNWs with a hole mobility on the order of $600cm^2/Vs$, approaching $1800cm^2/Vs$ that of bulk Ge.[4]

The high surface areas of nanotubes and nanowires suggest that surface effects could play dominant roles in determining their physical properties. The electrical properties of nanotube and nanowire devices could be affected by various surface species including molecules adsorbed from the environment.[14] For an example, hysteresis in the electrical characteristics of carbon nanotube FETs has been observed due to physisorbed water molecules on the nanotube surface and on the $SiO_2$ substrates around the nanotube.[15] Transistor prototypes based on several semiconducting nanowires including Si have been observed to exhibit hysteresis behavior.[16] The origin of hysteresis in nanowire FETs however, remains largely elusive thus far. To our knowledge, no systematic effort has been taken to elucidate or eliminate the origin of the hysteresis. Uncontrolled hysteresis is highly problematic since it will make the operations of electronic devices highly unpredictable. Removal of the hysteresis will require detailed understanding of the surface chemistry involved in various nanowires. This is particularly important and challenging for GeNWs due to the well-known poor surface oxide properties for Ge.

In this article, we present a systematic investigation of the electrical characteristics and surface chemical properties of p-type and n-type GeNWs. We first show that p- and n-type doping of GeNWs can be achieved by introducing desired doping gases during CVD growth of GeNWs. Detailed hysteresis behaviors of as-fabricated GeNW FETs are then described. A thermal annealing process is developed to eliminate or suppress such hysteresis. X-ray photoelectron spectroscopy is then used to elucidate the surface species for as-made and cleaned GeNWs. The spectroscopy results combined with electrical data lead to the conclusion that the hysteresis in GeNW FETs are caused by water molecules absorbed on GeNW surface oxides. Moreover, XPS measurements clearly reveal band bending in GeNWs due to Fermi level pinning of surface states. The importance of surface effects to nanowires is discussed especially for wires with sizes approaching the nanometer scale. We also present a preliminary surface passivation method based on atomic layer deposition of $HfO_2$ high-$\kappa$ dielectrics.

**Methods.**

**Materials.** Germane in helium ($GeH_4$, 10% in He), diborane in hydrogen ($B_2H_6$, 10ppm in $H_2$) and phosphine in hydrogen ($PH_3$, 150ppm in $H_2$) were purchased from Voltaix, Inc. (Branchburg, NJ). Gold nanoparticle colloid solutions were purchased from Sigma-Aldrich Co, and the diameters of the Au particles varied from 2nm to 20nm.

Transmission electron microscopy grids with Ni frames and thin $SiO_x$ supporting film were obtained from Structure Probe, Inc. (West Chester, PA).

**CVD growths of GeNWs.** (1) Synthesis of intrinsic GeNWs. Growth was carried out at 275°C with $GeH_4$ flow of 10 sccm (standard cubic centimeter per minute) and $H_2$ co-flow of 100 sccm in a 1" quartz tube furnace.[10] Typical growth time was 30-40 minutes. Catalyst was either Au nanoparticles (diameters = 2 to 20nm) deposited on $SiO_2$ or Si substrates from an aqueous solution of Au colloids (by soaking the substrates in a aminopropyltriethoxysilane APTES solution then in a Au colloid solution for 30 min) or a thin Au film (thickness ~3nm) evaporated from an electron-beam evaporator. (2) Synthesis of boron-doped p-type GeNWs. To grow p-type GeNWs, $H_2$ co-flow was replaced with a mixture of $H_2/B_2H_6$ ($B_2H_6$ concentration 10ppm). All other growth parameters were kept the same as for the intrinsic wires. (3) Synthesis of phosphorous-doped n-type GeNWs. Pure $H_2$ and $PH_3$ in $H_2$ ($PH_3$ concentration 150ppm) were used as co-flows ($H_2$ flow rate 80sccm and $H_2/PH_3$ flow rate 20sccm) of $GeH_4$ for n-type Ge nanowire growth. As shown later by transistor data, these growth conditions were effective in producing p- and n-doped GeNWs respectively. Our low 275°C CVD growth using germane was mild and highly reproducible. No significant reduction in the yield of GeNWs was observed due to the addition of $B_2H_6$ or $PH_3$. (4) Synthesis of GeNWs on TEM grids. Au nanoparticles were deposited on a TEM grid with $SiO_2$ supporting film in the same manner as on a Si or $SiO_2$ substrate. The grid was subjected to CVD growth and the resulting GeNWs were imaged by TEM directly after growth.[10]

**X-ray photoelectron spectroscopy (XPS).** XPS measurements were carried out on a PHI Quantum 2000 scanning microprobe equipped with an angle-resolved hemispherical electron energy analyzer and a base system pressure of $1 \times 10^{-9}$ torr. All measurements were performed using the Al $K_\alpha$ line with a photon-energy of 1486.6 eV. Charge compensation during acquisition was applied using the improved dual neutralization capability including an electron flood gun and an ion gun. C 1s (284.8 eV) on $SiO_2$ substrate were used to calibrate the process parameters and to ensure sufficient charge compensation. Ag and Cu standards were also employed to calibrate the system. The samples used for XPS studies were mats of p- or n-type GeNWs (random dense networks of GeNWs) grown on $SiO_2$ substrates with a thin Au film (3 nm) deposited on the substrates as catalyst.

**Transmission electron microscope (TEM) and scanning electron microscope (SEM).** The transmission electron microscope used was a Philips CM20 with a working acceleration voltage of 200 keV. The scanning electron microscope used was a FEI Siron field emission instrument operated at relatively low acceleration voltages of 1 keV to 5 keV.

**GeNW FETs with back-gate and $SiO_2$ gate-dielectrics.** $SiO_2$/Si ($p^{++}$) substrates were used for growth and integration of GeNWs into FET structures (Fig. 2a). The thickness of $SiO_2$ formed by dry oxidation was 10 nm (verified by ellipsometry) in the source (S), drain (D) and NW channel regions. The $SiO_2$ thickness was 500 nm in regions under large electrical bonding pads to avoid potential damages to the $SiO_2$ during electrical probing or wire-bonding. Details of the substrates design and preparation can be found in a previous work.[17] Au seeds (average particle size ~10nm) were deposited on the $SiO_2/Si$ surface in a patterned fashion[10,18] and used for CVD growth to produce GeNWs (p-type or n-type) from the patterned catalyst sites. After growth, samples with p-type NWs were annealed at 400 °C for 2 h intended for dopant activation and then soaked in 0.1 M HCl for 30 seconds to remove native surface oxides on the GeNWs. No annealing or activation was carried out for n-type GeNWs prior to device fabrication. Photolithography or electron-beam lithography patterning of photo-resist, metal evaporation (30 nm of Ti) and lift-off were then used to form S and D metal electrodes on top of the GeNWs. The length of the GeNW between the edges of the S and D electrode was typically L=3μm. FET operation was carried out by using the heavily doped Si as the back-gate electrode and the 10 nm $SiO_2$ as the gate dielectric (Fig.2a). Note that even though the GeNWs were treated in HCl (known to remove germanium oxides) prior to the fabrication steps, the GeNWs were exposed to the ambient environment for ~ 1-3 weeks through the device fabrication steps, which caused the formation of oxides on the GeNWs in the as-fabricated devices. Ti was used here for S and D electrodes to contact GeNWs since Ti-Ge contacts were found to remain intact after high temperature annealing up to 450°C. This stability was useful for our annealing experiments carried out for GeNW FETs, as described below.  Though optimization of contacts was not the main scope of the current work, we note that earlier, Pd was found to form more transparent contacts for GeNWs.[4]  Pd was not used here since the quality of the contacts tended to degrade upon annealing to 450°C. Also as a side note, we did not observe significant improvement of the Ti contacts with GeNWs upon annealing to 450°C.

**ALD of HfO$_2$ for top-gated GeNW FETs with high κ dielectrics.** ALD of HfO$_2$ films on the back-gated p-type GeNW FETs described above was carried out to produce top-gated devices with high k gate dielectrics (Fig. 9a). The back-gated p-type GeNW FETs were first 'cleaned' by annealing at 450°C for ~0.5 h in vacuum to remove surface oxides. The sample was then transferred to an ALD chamber for HfO$_2$ deposition after a brief and minimized exposure to air (<0.5 h). HfO$_2$ was deposited uniformly onto the sample (including GeNWs, S and D electrodes and the substrate) by reacting deionized water with tetrakis(diethylamido)hafnium (Hf[NEt$_2$]$_4$) precursor in a layer by layer fashion.[19] Nitrogen was used as the carrier gas, and the deposition temperature was 90°C. Each ALD reaction cycle involved alternating H$_2$O and Hf[NEt$_2$]$_4$ doses and the purge times were 350 s after the DI H$_2$O dose and 150 s after the Hf[NEt$_2$]$_4$ dose. The sample was exposed to ~ 80 cycles of ALD to deposit a HfO$_2$ film with nominal thickness of ~ 8 nm. After the dielectric deposition, a lithography step was carried out to pattern top-gate metal electrodes on top of the GeNW channels with HfO$_2$ sandwiched in between, forming top-gated GeNW FETs. Note that there was an under-lap between the gate and S/D electrodes (Fig.9a&9b) with the top-gate switching only the GeNW section directly underneath the top-gate. The Si back-gate was used to control the NW sections aside of the top-gate and was fixed at a constant voltage. Detailed operation of this type of transistor can be found in a previous pubication.[20]

**Annealing of GeNWs in vacuum for oxide removal.** Annealing was performed for both mats of GeNWs grown on $SiO_2$ substrates and GeNW FETs in a vacuum chamber with a base pressure of $5\times10^{-8}$ torr. The sample was placed on a metal stage equipped with a thermal heater and a thermal couple for temperature monitoring. Annealing was typically performed at 450ºC for 30 minutes, which was found to be efficient in volatilizing germanium oxides on GeNWs. After annealing, the GeNW FETs (p-type or n-type) were quickly transferred into a vacuum chamber ($5\times10^{-7}$ torr) for ex-situ electrical measurements, and the transfer process exposed GeNWs in the devices to ambient air for ~10-20 minutes. For XPS measurements, the GeNW mats were transferred into the XPS vacuum chamber after exposure to ambient air for a specified amount of time ranging from minutes to 1/2 h or to days.

**Results and Discussions.**

**Growth.** High yield of single crystalline GeNWs can be grown by our simple CVD method at 275°C with various Au seeds ranging from discrete Au colloidal nanoparticles to vapor-phase deposited thin Au films. The growth mechanism for the GeNWs involves the vapor-liquid-solid (VLS) process, as described previously.[10] Shown in Fig 1 is a low magnification TEM image of as-grown GeNWs from nominally 5 nm Au nanoparticles deposited on a TEM grid. The inset in Fig. 1 shows a lattice resolved high resolution TEM image of a ~ 4 nm diameter GeNW. We found that GeNW size was highly controllable by tuning catalyst sizes. The smallest wires we successfully and reproducibly grew were ~3 nm from ~2-3 nm Au nanoparticles. On 3 nm thick Au films, CVD growth led to the formation of a dense mat of GeNWs with diameters of 10-30 nm, suggesting that large Au particles were formed when the Au film was heated during CVD and these particles served as seeds for GeNWs growth.

**Characteristics of p-type GeNW FETs.** The back-gated FETs fabricated on B-doped GeNWs clearly exhibited p-type transistor behavior with about 3 orders of magnitude conductance (G) decrease under positive gate voltages (Fig.2c), indicating successful incorporation of p-type B-dopants into GeNWs during the CVD synthesis process. An immediate phenomenon we observed during electrical characterization of the un-passivated GeNW FETs (with NWs exposed to ambient air) was the large hysteresis in the current vs. gate voltage ($I_{ds}$-$V_{gs}$) curves. That is, the source-drain current is sensitive not only to gate voltage, but also to the gate voltage sweeping directions. The broken lines in Fig. 2c are $I_{ds}$-$V_{gs}$ curves recorded with a p-type GeNW FET under gate sweeping from $V_{gs} = –2V$ to 2V and then back to –2V at a sweep rate of $\Delta V_{gs}/\Delta t$=0.3 V/s. A ~ 1V difference in the threshold voltages was clearly present between the $I_{ds}$-$V_{gs}$ curves recorded under the two sweeping directions. The amount of hysteresis typically increased (i.e., larger differences in threshold voltages or larger hysteresis loops) under decreased gate sweeping rate $\Delta V_{gs}/\Delta t$ and was observed with all of the as-fabricated GeNW FETs (nearly 100 devices characterized) without exception. The hysteresis was nearly unchanged when the GeNW FETs were placed and measured in vacuum ($5\times10^{-7}$ Torr, Fig.2c), even under vacuum pumping for extended periods of time up to several days. At a first glance, the hysteresis behavior appeared to be a result of slow charging and discharging of states or chemical species on the GeNWs during gate sweeps.

**Elimination of hysteresis for p-type GeNW FET.** We carried out various treatments of the GeNWs aimed at eliminating and understanding the origin of the hysteresis. Our general strategy was to clean the GeNWs and thus eliminate the states or chemical species responsible for the hysteresis. After experimenting with various methods including solution phase chemical etching of Ge oxides,[21-23] we found that the most efficient method of eliminating hysteresis in the GeNW FETs was simple annealing in vacuum at ~ 450°C. The GeNW FETs were typically annealed in vacuum at ~450ºC for 1/2 h and then transferred to another vacuum chamber for ex-situ electrical measurements. During the transfer, the devices were briefly exposed to ambient air for ~20 min. After the brief exposure and when placed back in vacuum, the annealed p-type GeNW FETs exhibited almost no hysteresis (<30 mV difference in threshold voltage) as shown in Fig. 3a, and the result was independent of the gate voltage sweep rate in the range of $\Delta V_{gs}/\Delta t = 0.03$ to 0.6 V/s. These results illustrate that the thermal annealing step is effective in eliminating hysteresis for p-type GeNW FETs.

**Cause of Hysteresis.** The annealed GeNW FETs remained hysteresis free as long as kept in vacuum. Upon exposure of the devices to air, large hysteresis was observed immediately (within 1 min) as shown in Fig. 4a for a typical device. The hysteresis disappeared as soon as the sample was placed back into vacuum (Fig. 4b, circles) after a short (~ 10 minutes) exposure to air. The reversible hysteresis removal and re-appearance were observed for up to many vacuum/air-exposure cycles as long as the air exposure was brief (minutes). This result hinted that the hysteresis was caused by molecules adsorbed on the GeNW surfaces from the ambient air, and that for p-type GeNWs with surface oxides removed by 450°C annealing (see for example ref.21 and later sections for evidence of oxide removal), the molecules causing the hysteresis can be readily desorbed by pumping in vacuum.

In a control experiment, we exposed an annealed GeNW FET to dry air and observed no hysteresis at all in the device characteristics (Fig.4b, solid line). This hints that $H_2O$ molecules adsorbed on the GeNW surfaces are responsible for the hysteresis behavior of the GeNW FETs.

Long exposures of p-type GeNWs (after high temperature annealing/cleaning) to ambient air were found to cause hysteresis that were irreversible by vacuum pumping. For example, after 1-day exposure of an annealed p-GeNW FETs to ambient air, a hysteresis of ~200mV was observed (Fig. 4c) when measured with the device placed back in vacuum. After exposure to air for 1-week, the measured hysteresis in vacuum increased to ~500 mV (Fig.4d). This result was consistent with that the as-made p-type GeNW FETs (with the GeNWs exposed to air for typically > 1-week during device fabrication) exhibited large hysteresis that persisted in vacuum. As shown later, long air-exposure of p-type Ge leads to the formation of a surface $GeO_2$ layer that strongly adsorbs $H_2O$ molecules. The water molecules do not desorb from the $GeO_2$ layer on the GeNW in vacuum and are responsible for the irreversible hysteresis behavior of the transistors.

**Characteristic of n-type GeNW FETs and different hysteresis behaviors from p-type FETs.** N-type GeNW FETs were successfully fabricated with phosphorous-doped n-type GeNWs, as shown by the orders of magnitude conductance increase (vs. decrease for p-FET) under positive gate voltages (Fig.5a). This suggested successful incorporation of P-atoms into the GeNWs during CVD growth in the presence of $PH_3$. The as-fabricated n-type GeNW FETs (exposed to ambient air for ~ 1 week after growth) also exhibited large hysteresis (~1V) under back-and-forth gate voltage sweeps (Fig.5a). Vacuum annealing experiments at 450°C were conducted on n-type GeNW FETs, followed by transferring of the devices to another vacuum chamber for electrical characterization, during which the devices were exposed to ambient air briefly (20 min). This treatment led to clear reduction of hysteresis, but unlike the p-type GeNW FETs, small hysteresis persisted for the n-type FETs (Fig.5b). This result, as shown later by XPS data, is due to rapid $GeO_2$ formation on n-type Ge (which differs from p-type Ge) during the brief air-exposure sample-transfer step, causing strong water adsorption on the $GeO_2$ on the nanowires and therefore the irreversible hysteresis. After annealing of the n-type GeNW FETs, exposure of the devices to air for ~ 1 week led to large hysteresis similar to that of as-fabricated n-FETs (Fig.5c).

**XPS of p-type and n-type GeNWs with Native Oxides.** XPS experiments were performed on mats of GeNWs to investigate the species on the nanowire surfaces and then correlate the results with electrical properties of the GeNW transistors. Ge is known to oxidize in various environments and form GeO and/or $GeO_2$ depending on the environmental conditions.[24,25] The freshly grown GeNWs contained large numbers of surface atoms and are therefore highly reactive. Once synthesized, the GeNWs were retrieved from the CVD growth chamber and exposed to the ambient air. For both p- and n-type GeNWs exposed to air for >~1 day after CVD synthesis, XPS revealed the presence of 'native' GeO and $GeO_2$ on the GeNW surfaces, as shown in the Ge 3d and Ge 2p spectra in Fig.6. The zero valent $Ge^0$ states are peaked in the 3d spectra around 29.8eV to 29.95eV (Fig. 6a and b) and the germanium oxides are around 32.7eV to 33eV. Curve fittings (Fig.6a and b) reveal that the oxide peaks correspond to combinations of $Ge^{2+}$ (GeO) and $Ge^{4+}$ ($GeO_2$). The peak positions of $Ge^0$ and $Ge^{2+}$ and $Ge^{4+}$ are consistent with literature values reported for Ge wafers, with ~0.75 eV shift per oxidization state.[22,25,26] The presence of surface native oxides can also be clearly seen from Ge 2p spectra in Fig. 6c and 6d for p and n-type GeNWs respectively.

**XPS of p-type and n-type GeNWs after annealing at 450°C.** It is known that both GeO and $GeO_2$ can be removed from Ge surfaces by thermal annealing in vacuum. GeO starts to vaporize above ~ 400°C. $GeO_2$ by itself is stable up to higher temperatures,[27] but when in contact with Ge, $GeO_2$ reacts with Ge at >~400 C, converts to GeO and then sublimes.[21] Therefore, relatively clean and oxide-free surfaces are expected for GeNWs after heating above 400°C. Our XPS data recorded with both p- and n-type GeNWs after annealed to 450°C indeed revealed the removal of the oxides signaled by the significantly reduced peaks corresponding to the oxides (Fig.7a and b).

**Different oxidation behaviors of p- and n-type GeNWs.** The XPS data of p-and n-GeNWs after annealing shown in Fig. 7a and b still displays weak signals in the region corresponding to Ge oxides due to sample transfer and exposure to air for ~2 min after the annealing. We have systematically varied the air-exposure time and recorded XPS spectra to investigate the oxide formation processes on p- and n-type GeNWs. For p-type GeNWs, after annealing and 2 min air-exposure, a shoulder-like feature can be seen near the $Ge^0$ peak in the Ge 3d spectra, and peak fitting reveals that the shoulder mainly corresponds to GeO. Longer air exposures (1/2 h to 1day) cause the shoulder to grow and shift towards higher binding energies (Fig.7c), and the spectrum recorded after 1-day exposure shows a clear peak corresponding to $GeO_2$ (Fig.7c). Curve fitting also shows GeO in coexistence with $GeO_2$. These results suggest that upon exposure of 'clean' p-type GeNWs to air, oxide quickly forms on the Ge surfaces and the oxide species is predominantly GeO. Longer air exposure causes the growth of oxides on the surfaces and the formation of appreciable amount of $GeO_2$ on p-type GeNWs is relatively slow and takes a relatively long time (hours).

The n-type GeNWs exhibit different oxidation behavior than the p-type wires. After annealing and a brief 2 min air-exposure, appreciable amount of $GeO_2$ is already detected on the nanowire surfaces (Fig.7b). Longer exposures (1/2 h to 1day) of the n-GeNWs to air cause the intensity of the oxide peak to grow, but unlike the p-type GeNWs, no significant shift in the peak position is observed (Fig. 7d). These results suggest that $GeO_2$ formation is rapid on n-type GeNW surfaces, within minutes of exposure to air. Longer exposure times in air lead to thicker $GeO_2$ films.

It is to our knowledge the first time that dopant-dependent reactivity is reported for Ge. The p-type GeNW oxidation behavior, i.e., rapid GeO formation followed by slow $GeO_2$ growth is similar to previous reports for p-type Ge wafers.[24] No systematic investigation on n-type Ge wafer oxidation has been reported previously. Note that dopant dependent chemical reactivity is known for semiconductors including Si. An example is the drastically different etching rates for p- and n-type Si wafers in $XeF_2$ etching gases.[28] Nevertheless, a detailed understanding of the different oxidation behaviors of n- and p-type semiconductors is currently lacking and we leave it for future investigations.

**Band bending in oxidized GeNWs due to interface states.** In addition to the oxide species on GeNWs, our XPS results also revealed band bending caused by Ge-$GeO_x$ interface states on GeNWs with surface oxides. For p-type GeNWs, the removal of the native oxides by annealing at 450°C leads to a –0.3 eV shift of the $Ge^0$ 3d peak from 29.8eV to 29.5eV (Fig.7a). For n-type GeNWs, an opposite +0.1 eV shift from 29.95eV to 30.05eV is observed after oxide removal (Fig.7b). These results suggest the existence of Ge/$GeO_x$ interface states residing within the band gap of Ge, giving rise to Fermi level pinning at the interface and band bending away from the interface with a characteristic screening length $d$ (Fig.8). The band bending causes an increase (for p-type, Fig. 8a) or decrease (for n-type, Fig. 8b) in the distance between the Ge 3d core-level to the Fermi level, and thus increases or decreases the binding energy of the Ge 3d level relative to bulk Ge for p- and n-type GeNWs respectively. Oxide removal at the surface eliminates the interface states and band bending, thus shifting the Ge peaks towards that of bulk Ge. The band-bending behaviors reported here for p- and n-type GeNWs are similar to those previously observed with p- and n-type Ge wafers.[22,29] Band bending due to surface interface states has also been reported for other semiconductors including Si and GaAs.[30,31] [32] Note that for oxide-free p- and n- type GeNWs, the $Ge^0$ 3d peaks are at 29.5eV and 30.05 eV respectively (Fig. 7a and b.), and the 0.55 eV difference approximately matches the Ge band gap of 0.6 eV. The $Ge^0$ 3d peak shift for n-GeNW is relatively small after oxide removal compared to the p-type, suggesting the interface states are close to the conduction band in n-type GeNWs with native oxides.

**Correlating hysteresis of GeNW FETs with surface XPS data.** Correlating our electrical and XPS data suggest that surface bound water is responsible for hysteresis in GeNW FETs. We can also infer that water association with $GeO_2$ on GeNW surfaces is strong and does not desorb by simple vacuum pumping, while the opposite appears to be true for water-GeO interaction. These are consistent with that as-fabricated p- and n-type GeNW FETs exhibit hysteresis that cannot be removed by placing the devices in vacuum. For p-type GeNW FETs, hysteresis is diminished after annealing in vacuum even with a brief re-exposure to air. This is consistent with XPS data of $GeO_2$ removal by vacuum annealing and predominantly GeO formation (rather than $GeO_2$) during the brief air re-exposure. For n-type GeNW FETs, hysteresis persisted after oxide removal followed by brief air exposure due to rapid formation of $GeO_2$ and strong association with water molecules. It is therefore rather difficult to eliminate surface bound water and related hysteresis in n-type Ge devices if any short air-exposure of clean n-type Ge surfaces can not be avoided.

The interface states due to Ge surface oxides could also act as charge traps and caused hysteresis behavior to the electrical characteristics of FETs. However, for our p-GeNW FETs exposed to pure oxygen or dry-air after vacuum annealing, no significant hysteresis was observed (Fig.4b) even after extended (2 h) exposure to pure oxygen, while XPS data revealed the formation of GeO and $GeO_2$ on p-type Ge during such exposure.[24] Hysteresis was observed immediately once the devices were taken out of the pure oxygen and exposed to ambient air (Fig.4a). These results confirm $H_2O$ molecules on the oxide surface rather than the Ge/$GeO_x$ interface states as the main cause to the type of hysteresis concerned in the current work. The hysteresis is large (on the order of ~ 1V), slow and dependent on gate sweep rate (in the range of $\Delta V_{gs}/\Delta t = 0.03$ to 0.6 V/s). It is possible that $GeO_x$/Ge interface states also give rise to hysteresis behavior, but at a different time scale than the one probed here by DC electrical measurements.

**Water on Ge oxide surfaces.** XPS data clearly revealed that GeNWs exposed to air contain native oxides on surfaces. The suggested strong interaction/association between $H_2O$ from the ambient air and surface $GeO_2$ is reasonable since $GeO_2$ is known to be soluble in $H_2O$. The surface chemistry of $SiO_2$ and Si is much better studied than for the Ge system. The existence of physisorbed water layers on native $SiO_2$ on Si is well known and in fact, approximately one monolayer of $H_2O$ is hydrogen bonded to the hydroxyl groups on $SiO_2$. As a result of the strong binding, desorption of the final surface water layer on $SiO_2$ only occurs above ~ 250°C.[33] While systematic spectroscopy work is needed to elucidate detailed $H_2O$ interactions with $GeO_x$/Ge, we note that several previous infrared spectroscopy work on $GeO_x$/Ge surfaces have indeed identified adsorbed water species.[34-35] A temperature programmed desorption (TPD) experiment revealed the existence of water and related species on Ge surfaces at high temperatures up to 250-300 °C before desorption during heating.[36]

**Water in various bottom-up electronic devices including nanotubes, nanowires and organic FETs.** For modern Si electronics, much is known about the causes and elimination of hysteresis in metal-oxide semiconductor (MOS) FETs. Slow states and hysteresis related to water has been reported for MOSFETs without proper encapsulation/passivation, as well as for MOS capacitances and tunnel diodes with hydrated $SiO_2$ gate insulators.[37-39] Recently, we have reported hysteresis in single-walled carbon nanotube FETs formed on $SiO_2$ substrates due to water molecules physisorbed on nanotube surfaces and on $SiO_2$ surrounding the nanotube.[15] The current GeNW results represent another example of hysteresis phenomena in nanomaterials-based devices due to surface water. The water effect appears to be general in various un-passivated FETs based on materials derived by bottom-up chemical methods. There are several reports of large hysteresis observed in other types of nanowire FETs and in various organic transistors built on $SiO_2$ substrates.[40,41] We believe that water is a likely cause of hysteresis in these systems as well. The removal of surface bound water will be dependent on the specific systems and surface chemistry involved. For instance, in the case of carbon nanotube FETs, coating the nanotubes with a polymer, polymethylmathacrylate (PMMA) followed by baking can remove water adsorbed on the nanotube and on $SiO_2$ surfaces.[15] Hydrogen bonding of PMMA with the latter is responsible for removing the water layer on $SiO_2$.[42] In the case of GeNWs, similar PMMA passivation method was unsuccessful in removing the hysteresis, suggesting that $H_2O$ association with surface $GeO_2$ is strong and cannot be removed by the PMMA treatment. Note that it is possible that $H_2O$-derived surface species such as –OH groups may contribute to hysteresis. Systematic infrared spectroscopy work is indeed needed to clearly identify the species on Ge surfaces treated under various conditions, and to correlate with the electrical properties of Ge devices.

**Mechanism of hysteresis due to water.** The detailed mechanism of surface water causing hysteresis behavior in un-passivated electronic devices is not fully understood and requires further investigation. We suggest that a possible mechanism is due to the large electric dipole of water molecules. In an electric field on the order of ~1V/nm generated by the gate voltage, water molecules will respond to the field and be oriented. The water molecule orienting and relaxation processes under sweeping gate voltages and electric fields could exert varying dipole fields to the channel of the FETs, giving rise to additional gating mechanism to the transistors and therefore hysteresis. Other mechanisms are certainly possible including the relaxation of mobile impurities and ions on the surface of the transistor materials in the presence of water.

**Doping levels and radial screening lengths in NWs.** The doping level ($n$) in GeNWs can be estimated based on[43] resistivity (ρ) of the nearly hysteresis-free GeNW FETs (Fig. 3 and 5b for p- and n-type respectively) at zero gate-voltage

$$n = \frac{1}{e\mu\rho} \qquad (1)$$

where µ is carrier mobility estimated from the FET characteristics (e.g., Fig.3b)

$$\mu = \frac{dI_{ds}}{dV_{gs}} \times \frac{L^2}{C_{ox}} \times \frac{1}{V_{ds}} \qquad (2)$$

The gate-capacitance $C_{ox}$=3.8×10$^{-16}$F is estimated from

$$C_{ox} = \frac{2\pi\varepsilon_0\varepsilon L}{\cosh^{-1}(\frac{R+t_{ox}}{R})} \qquad (3)$$

where t$_{ox}$=10nm is the SiO$_2$ gate oxide (ε~4) thickness, L ~ 3 µm is the length of the GeNW between source and drain electrodes and R ~ 5 nm is the NW diameter. The resistivity ρ of GeNW under zero gate voltage is determined from

$$\rho = \frac{dV_{ds}}{dI_{ds}} \times \frac{L}{A} \qquad (4)$$

where *A* is the NW cross-sectional area. Through this analysis, we obtained a B-doping level of *n*=3×10$^{18}$/cm$^3$ for p-type GeNWs and a P-doping level of *n*=6×10$^{18}$/cm$^3$ for n-type GeNWs. Note that the ratio of B:Ge in the feeding gases for GeNW CVD growth was 1:500 and 1:300 for P:Ge, corresponding to maximum possible doping concentrations of 8×10$^{19}$cm$^{-3}$ and 1.2×10$^{20}$cm$^{-3}$ respectively for p and n-type GeNWs.

The extent of band bending due to Fermi level pinning by surface states can be estimated for nanowires. For a planar surface with surface Fermi level pinning potential of $\phi_0$, the screening length (or band bending length extending into the semiconductor with doping concentration of $n$) $d$ is given by[44]

$$d = \sqrt{\frac{4\pi\varepsilon_0\varepsilon\phi_0}{2\pi n}} \qquad (5)$$

where $\varepsilon_0$ is the permittivity of vacuum and $\varepsilon$ is the dielectric constant of the semiconductor ($\varepsilon \sim 16$ for Ge). For a nanowire with radius R, the depletion length $d$ (along the radial direction of the NW, Fig. 8c) can be obtained after solving the Poisson's equation using cylindrical coordinates (for $d<R$),

$$-\frac{4\pi\varepsilon_0\varepsilon\phi_0}{2\pi n} = R^2\left(1-\frac{d}{R}\right)^2 \times \left[1-2\ln\left(1-\frac{d}{R}\right)\right] - R^2 \quad (6)$$

For $d<<R$, to the first-order of the NW curvature ($d$/R), the radial screening length $d$ in the NW is given by,

$$\frac{4\pi\varepsilon_0\varepsilon\phi_0}{2\pi n} = \left(1-\frac{d}{3R}\right)d^2 \qquad (7)$$

 which suggests that the screening length in a nanowire is longer than that in the planar surface case. This is a reasonable result since screening becomes less effective when approaching the core of a NW due to the reduced radial dimension. Note that Eq. 6 and 7 are valid only for $d<R$.  For $d\geq R$, the whole nanowire along the radius will be depleted and it is no longer meaningful to calculate $d$.

If we assume that the GeNWs are large in radius with $R>>d$, then for our p-type GeNWs with $n=3\times10^{18}$cm$^{-3}$ and Fermi level pinning of $\phi_0=0.3$eV (Fig. 7a), a screen length of $d\sim13$nm is estimated based on Eq. 5. For n-type GeNWs with $n=6\times10^{18}$cm$^{-3}$ and $\phi_0=0.1$eV, $d\sim6$nm is obtained. These lengths represent the lower limit of the true screening lengths for the finite radius nanowires. The fact that they are similar to our GeNWs radii in the range of R=5 to 15 nm suggests that for the GeNWs with native oxides, Fermi level pinning and band bending at the surface will affect a large part of the NW 'bulk'. Such a surface effect could significantly alter the electrical properties of the wires. For p-type NWs, for instance, removal of surface oxides leads to a shift of threshold voltage by +2 V in the transistor characteristics (Fig. 2c vs. Fig.3a), indicating that the NWs with surface oxides are un-intentionally 'n-doped' as a result of band bending extending deep inside the wires from the surface due to surface states Fermi level pinning (Fig. 8a).

The analysis above points to the importance of surface to the bulk properties of semiconducting nanowires. Clearly, as the size of NW decreases, the dominance of surface effect will increase. Surface passivation is particularly important to GeNWs due to the high density of surface states ($>10^{14}/cm^2$).[45] Surface effects could also become significant in nanowires of Si or other types of semiconductors with low surface-state densities ($\sim 10^{12}/cm^2$)[45] especially if the diameters of the wires become very small. These issues will need to be systematically addressed by future research.

**Preliminary result of passivation of GeNW FETs by ALD of HfO$_2$.** We carried out preliminary experiments for passivating GeNW FETs. For p-type GeNW FETs, we first removed surface oxides on the GeNWs by vacuum annealing at 450°C and then carried out ALD HfO$_2$ at 90°C with minimum exposure of the devices to air after the annealing ($\sim$ 10 min total). Top-gate GeNW FETs obtained this way exhibited negligible hysteresis under top-gating when measured in vacuum or dry environments. These devices can be stored in ambient air for days without exhibiting large hysteresis when measured in a dry environment. This result suggests that the high-κ films on p-type GeNW FETs can be used as the first layer of passivation while serving as the gate insulator. Further work will be needed to investigate the nature of the $HfO_2$/Ge interface and understand the passivation effect. It will also be necessary to devise further passivation layers on top of the high κ films to completely block water, oxygen and other molecular species from the ambient air.

**Conclusions**

Single crystalline germanium nanowires with p- and n-dopants were synthesized by chemical vapor deposition at 275°C and used to construct p- and n-type field effect transistors respectively. The growth condition is mild and the size of the nanowires can be well controlled by the size of the Au nanoparticle seeds down to the 3-5 nm scale. Electrical transport and XPS measurements were carried out to understand the Ge surface chemistry and its role in the electrical characteristics of GeNWs. Large hysteresis observed in the electrical properties of non-passivated GeNWs was found to be mainly due to water molecules strongly bound to $GeO_2$ on GeNWs. Water adsorption on GeO was found to be weaker than on $GeO_2$ and can be easily removed. N-type GeNW FETs were more susceptible to hysteresis behavior due to more reactive surfaces towards rapid $GeO_2$ formation than p-type Ge. P-type GeNWs on the other hand, form GeO on the surfaces initially when exposed to air and then appreciable amount of $GeO_2$ after long exposures (hours). This is the first time that different chemical reactivity and hysteresis characteristics are observed for p- and n-doped Ge.

Opposite band bending due to interface states between Ge and surface oxides were observed for p- and n-type GeNWs. An annealing method was devised to remove surface oxides on GeNWs and eliminate or suppress hysteresis for GeNW FETs. Further, it was found that thin high-κ dielectric $HfO_2$ films grown on $GeO_2$-free GeNW surfaces by atomic layer deposition (ALD) was effective in serving as the first layer of surface passivation for GeNWs.

The current work reveals fundamental chemical and physical properties of Ge and may have important implications to Ge as a candidate material for future electronics. Further, the results should be of interest beyond Ge since surface water appears to be a generic cause to hysteresis in electronic devices based bottom-up chemically derived nanomaterials ranging from carbon nanotubes to various semiconductor nanowires or even organic transistors. Lastly, the depletion length or screening length along the radial direction of nanowires is evaluated. The result suggests that surface effects could be dominant over the 'bulk' properties of small diameter wires.

## Acknowledgements

We are grateful to Prof. S. Bent and R. Chen for discussions and sharing with us unpublished IR data. This work is supported by Stanford INMP, MARCO-MSD,

SRC/AMD, a Packard Fellowship, Sloan Fellowship and a Dreyfus Teacher-Scholar.

**Figure Captions**:

**Figure 1**. A TEM image of GeNWs grown from ~ 5nm Au nanoparticles. Inset: A high resolution TEM image of a GeNW with diameter ~ 4nm.

**Figure 2**. Back-gated p-type GeNW FETs. (a) A schematic view of the device structure. The $SiO_2$ thickness is $t_{ox}$=10 nm. The doped Si substrate is used for back-gate. (b) A SEM top-view image of an individual GeNW device recorded at an acceleration voltage of 5 keV. The electrodes and GeNW appear dark under this imaging condition sine they conduct electrons better than the $SiO_2$ background. (c) A typical as-fabricated un-passivated GeNW FET exhibiting large hysteresis under back and forth gate sweeping (direction marked by arrows, $V_{ds}$=100mV) measured with the device placed in air and in vacuum ($5\times10^{-7}$ Torr) respectively.

**Figure 3.** p-type GeNW FETs characteristics after removal of surface oxide by annealing in vacuum at ~450°C. (a) Current vs. gate voltage ($I_{ds}$-$V_{gs}$) curves showing nearly eliminated hysteresis after annealing (measured in vacuum) of the same device as in Fig. 2. The bias voltage used to record this curve was $V_{ds}$=100 mV. (b) Current vs. voltage ($I_{ds}$-$V_{ds}$) curves under various gate voltages as labeled.

**Figure 4.** p-type GeNW FET hysteresis vs. various environment. $V_{ds}$=100mV for all measurements. (a) Large hysteresis in the $I_{ds}$-$V_{gs}$ characteristics of the device in Fig.3 observed during a brief exposure to ambient air after annealing and surface oxide removal. (b) Hysteresis diminished when the device was placed in vacuum (symbols) and dry air (solid line). (c) After 24 h exposure of the device to ambient air, the device exhibited hysteresis that persisted when measured in vacuum. (d) After one-week exposure of the device to air, the device exhibited larger hysteresis when placed back into vacuum than in (c).

**Figure 5.** Hysteresis of n-type GeNW FETs. (a) An as-fabricated n-type GeNW FET exhibited large hysteresis with the device placed in vacuum. (b) The device exhibited significantly reduced hysteresis after 450ºC annealing in vacuum, a brief exposure to air and placed back in vacuum. (c) After one-week exposure to ambient air, large hysteresis recovered to the device. $V_{ds}$=100mV for all measurements shown here.

**Figure 6.** XPS spectra of p-type and n-type GeNWs with native oxides. (a) and (c) correspond to p-type GeNWs for Ge 3d and 2p spectra respectively. The inset in (a) shows an SEM image of the GeNW mat sample used for XPS experiments. The labeled peaks are curve fitting results for the experimental data, showing various surface chemical compositions. (b) and (d) are for n-type GeNWs.

**Figure 7.** XPS spectra of p-type and n-type GeNWs before and after 450°C annealing in vacuum. (a) p-type GeNWs 3d spectra before (red) and after (blue) annealing. A brief exposure to air was involved during sample transfer to XPS chamber. (b) Similar data to (a) for n-type GeNWs. (c) Evolution of XPS spectrum for p-type GeNWs after vacuum annealing and exposure to air for various periods of times as indicated. Note that the Ge 3d peak exhibited only slight shifts after exposure to air for up to 1 day. The degree of band bending was smaller than the sample in (a) that had been exposed in air for > 1 week due to less oxidation. (d) Similar data to (c) for n-type GeNWs.

**Figure 8.** Schematic illustration of opposite band bending for (a) p- and (b) n-type GeNWs respectively due to Fermi level pinning of surface states. (c) shows the depletion layer (thickness = $d$) of a GeNW due to band bending caused by Fermi level pinning at the surface.

**Figure 9.** Top-gated p-type GeNW FETs with high-κ $HfO_2$ dielectrics. (a) A schematic side view of the device. (b) A top-view SEM image of the device. SEM acceleration voltage=5keV. (c) $I_{ds}$-$V_{gs}$ characteristics of a typical device exhibiting little hysteresis ($V_{ds}$=100mV).

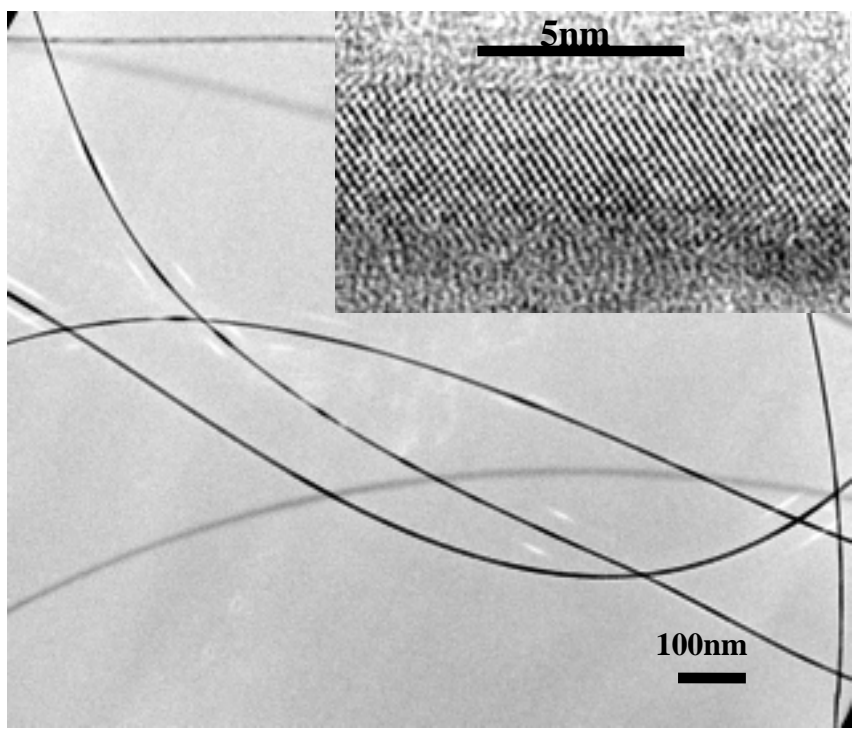

**Figure 1**

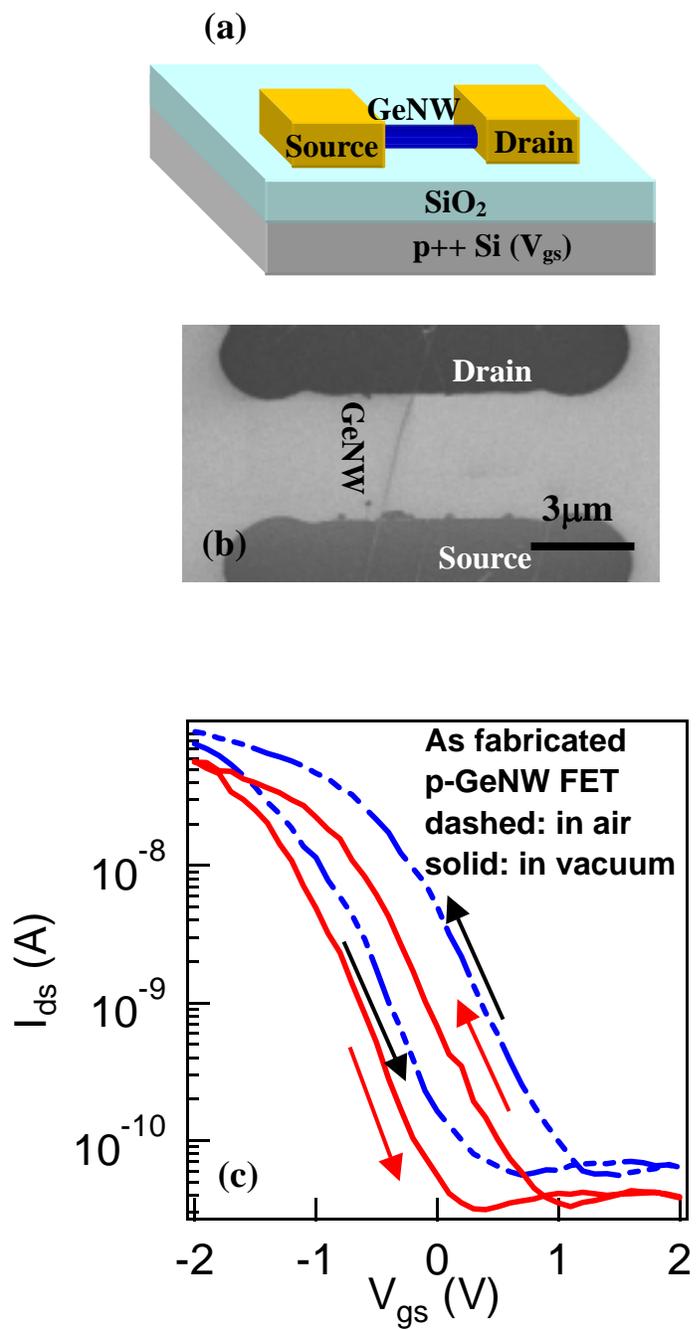

(a)

GeNW
Source    Drain
SiO₂
p++ Si (V_gs)

(b)

Drain
GeNW
3μm
Source

(c)

As fabricated
p-GeNW FET
dashed: in air
solid: in vacuum

$I_{ds}$ (A)

$V_{gs}$ (V)

**Figure 2**

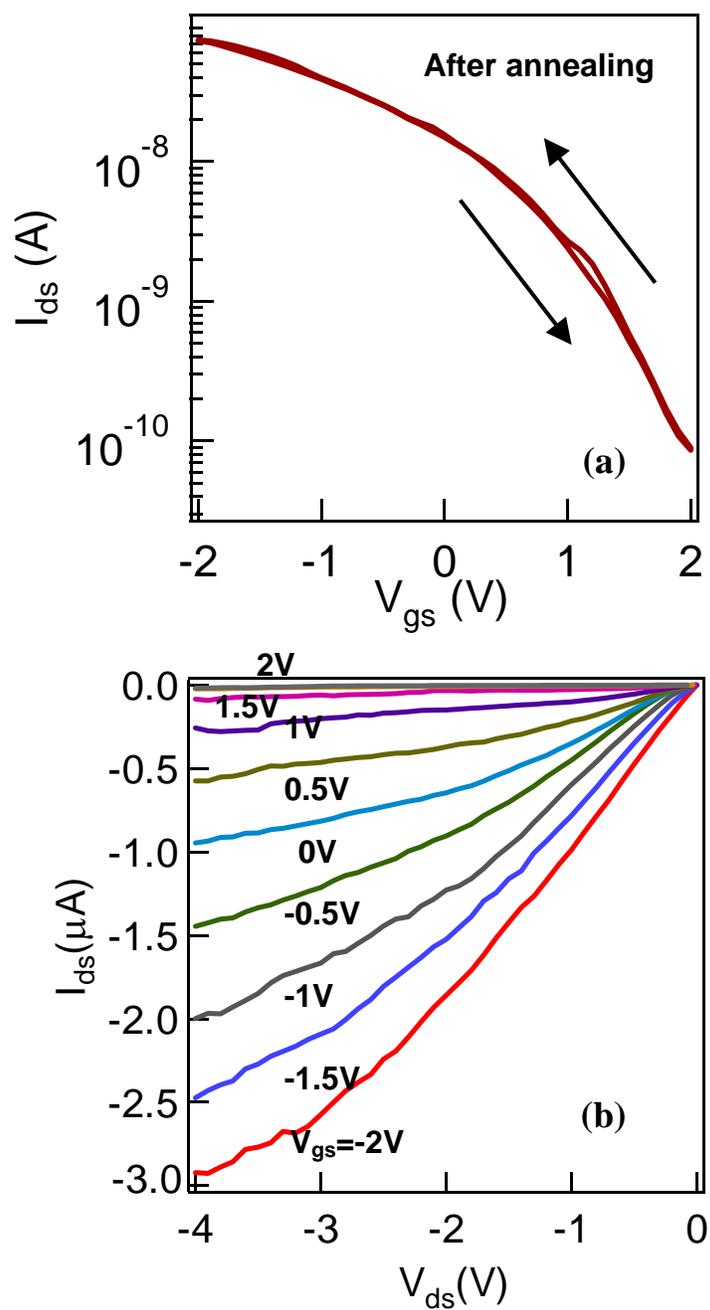

**Figure 3**

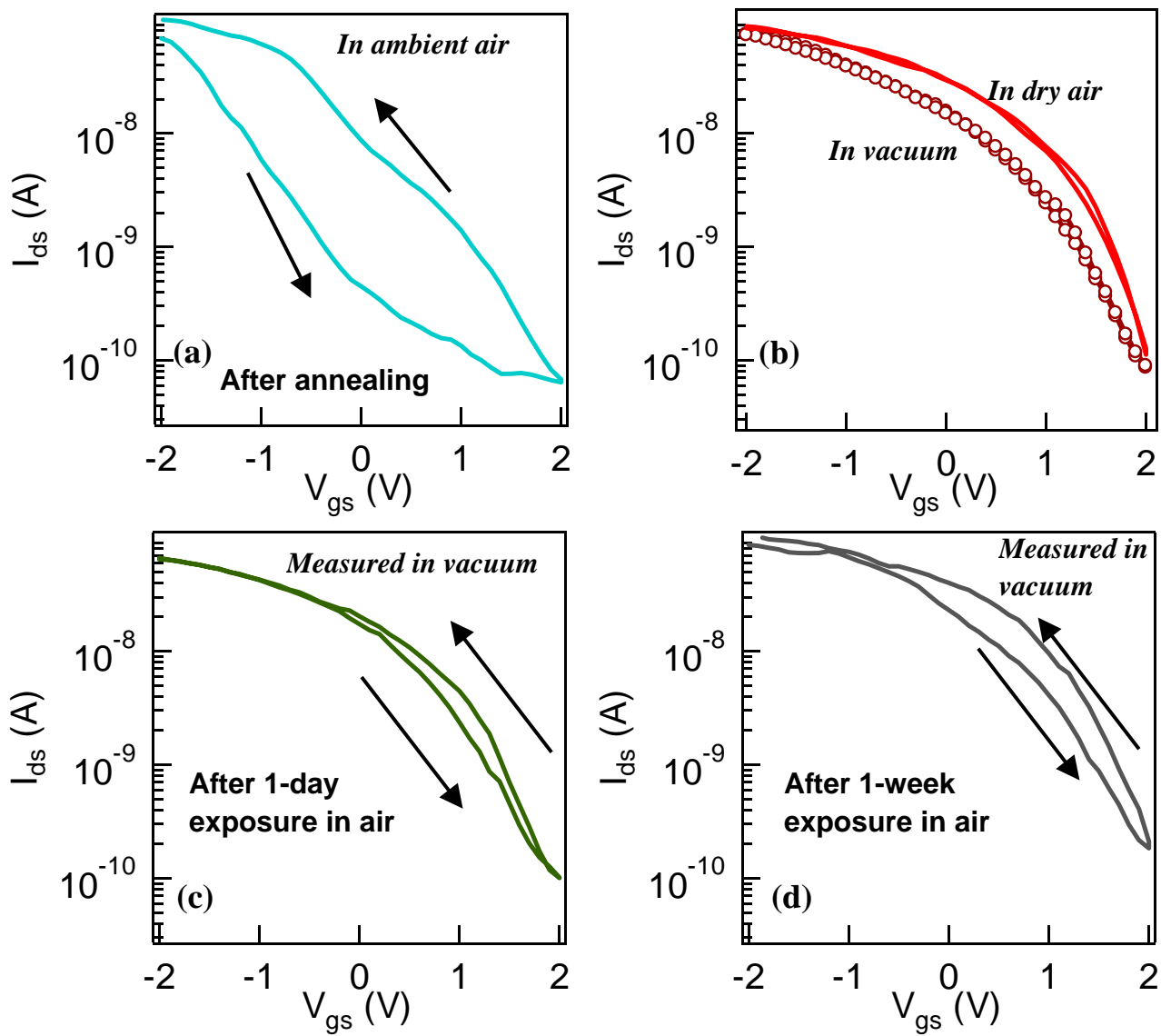

**Figure 4**

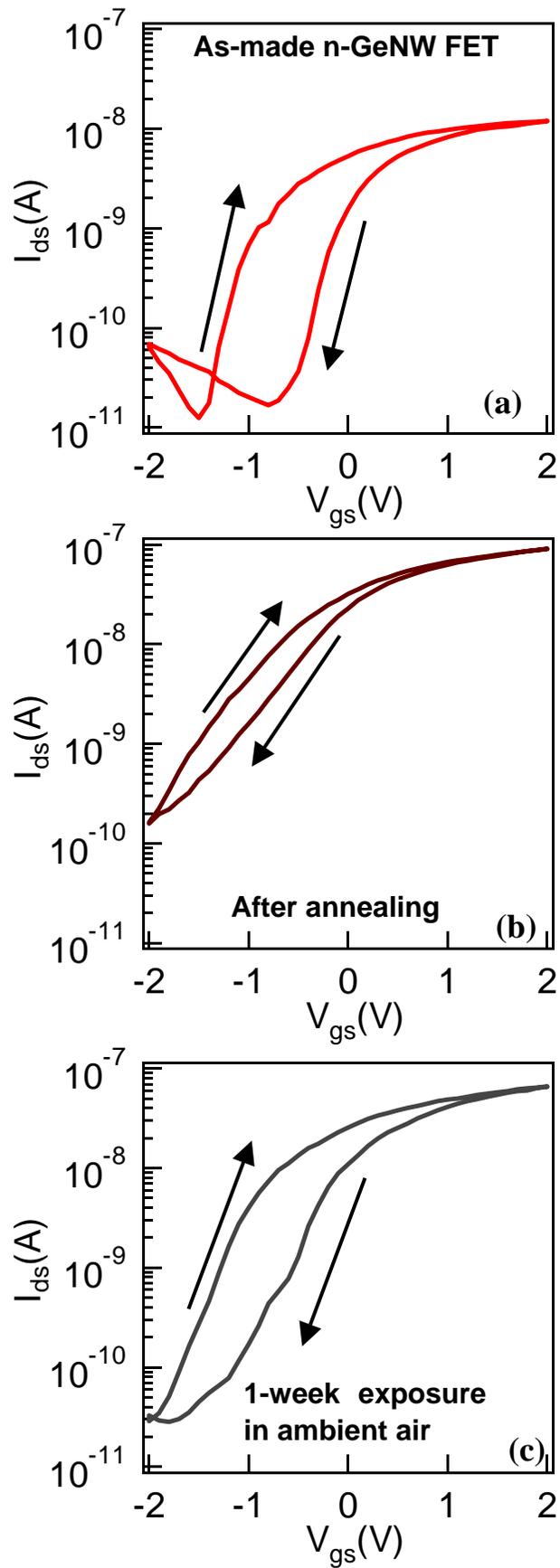

**Figure 5**

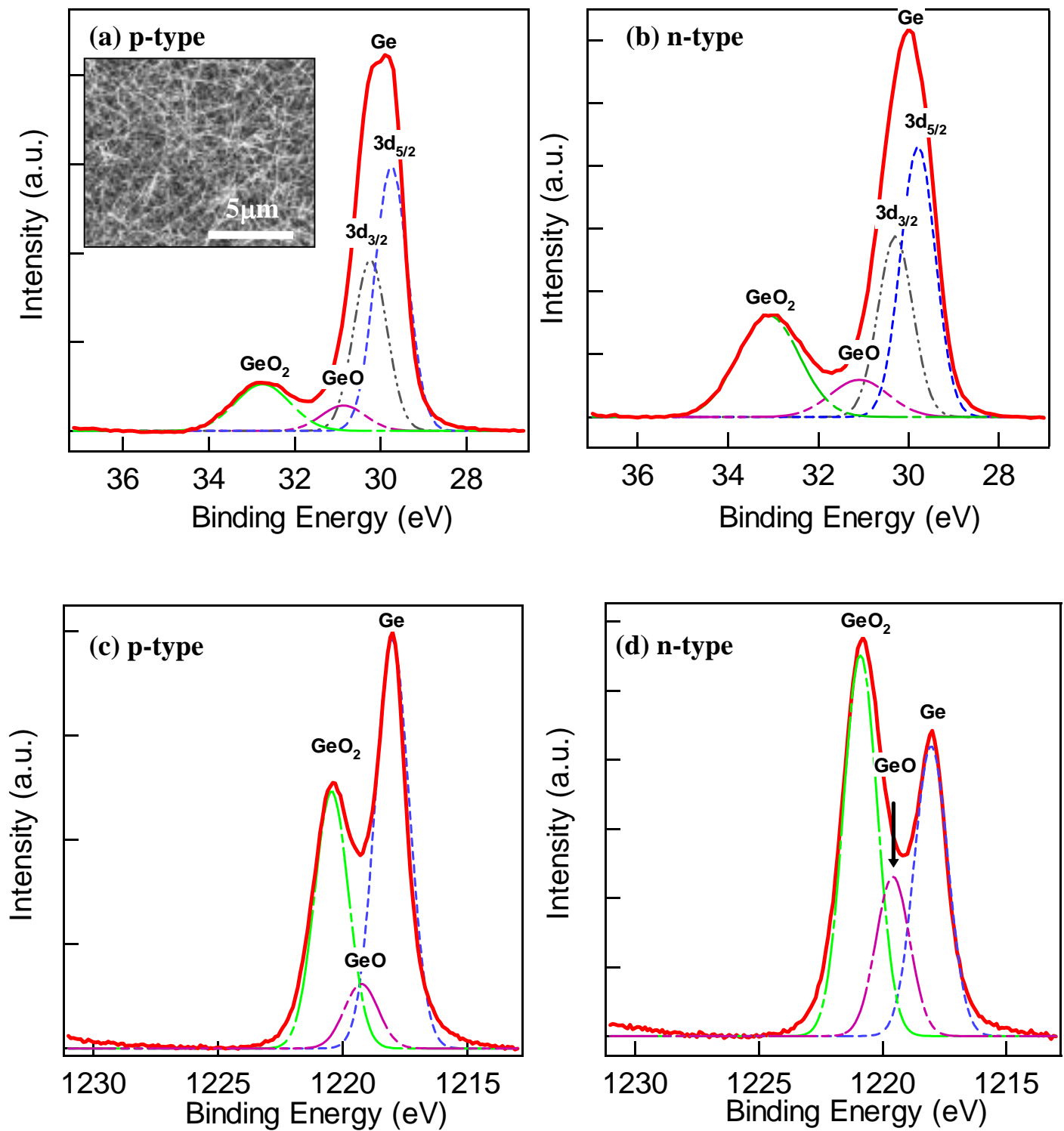

**Figure 6**

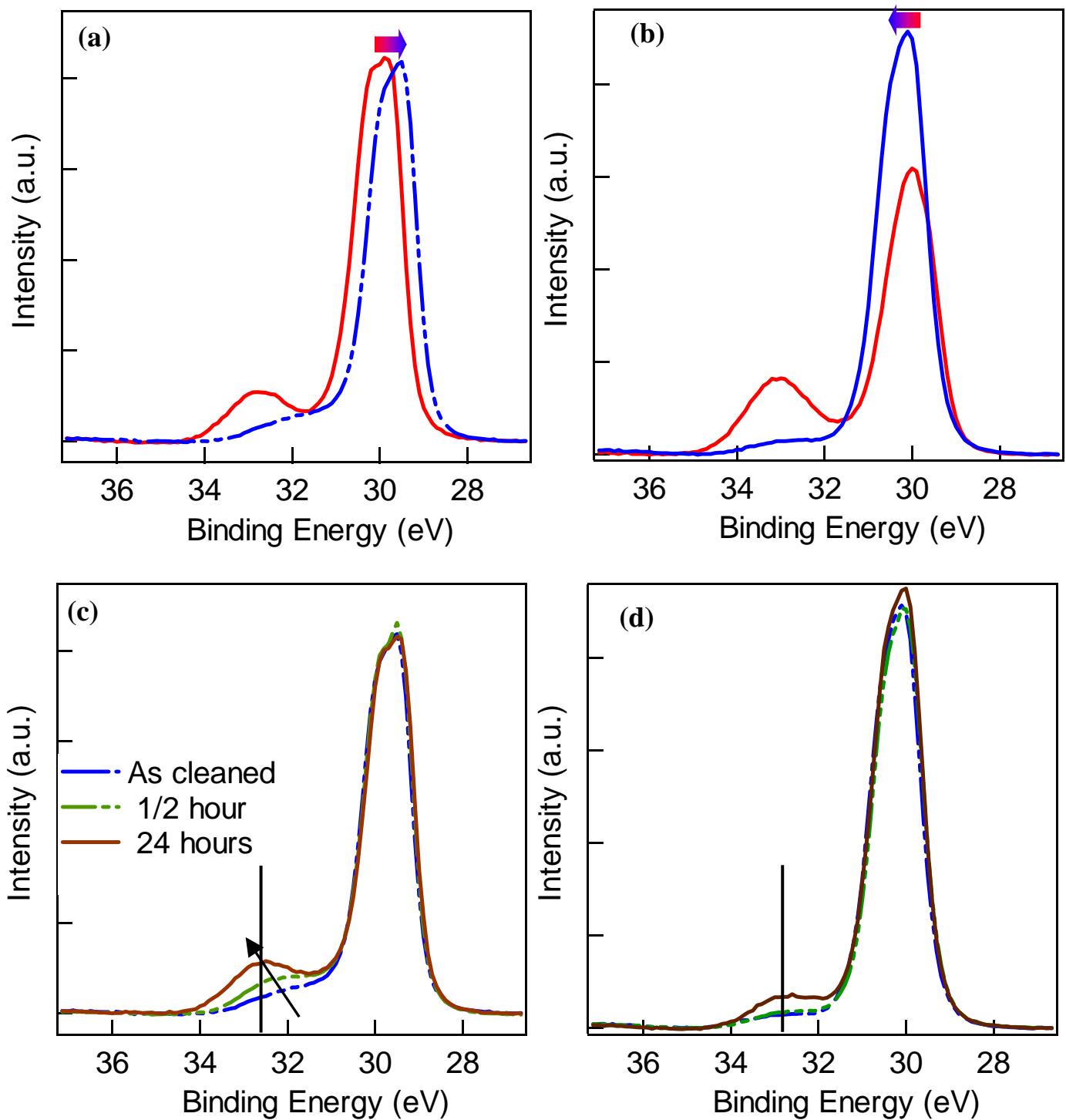

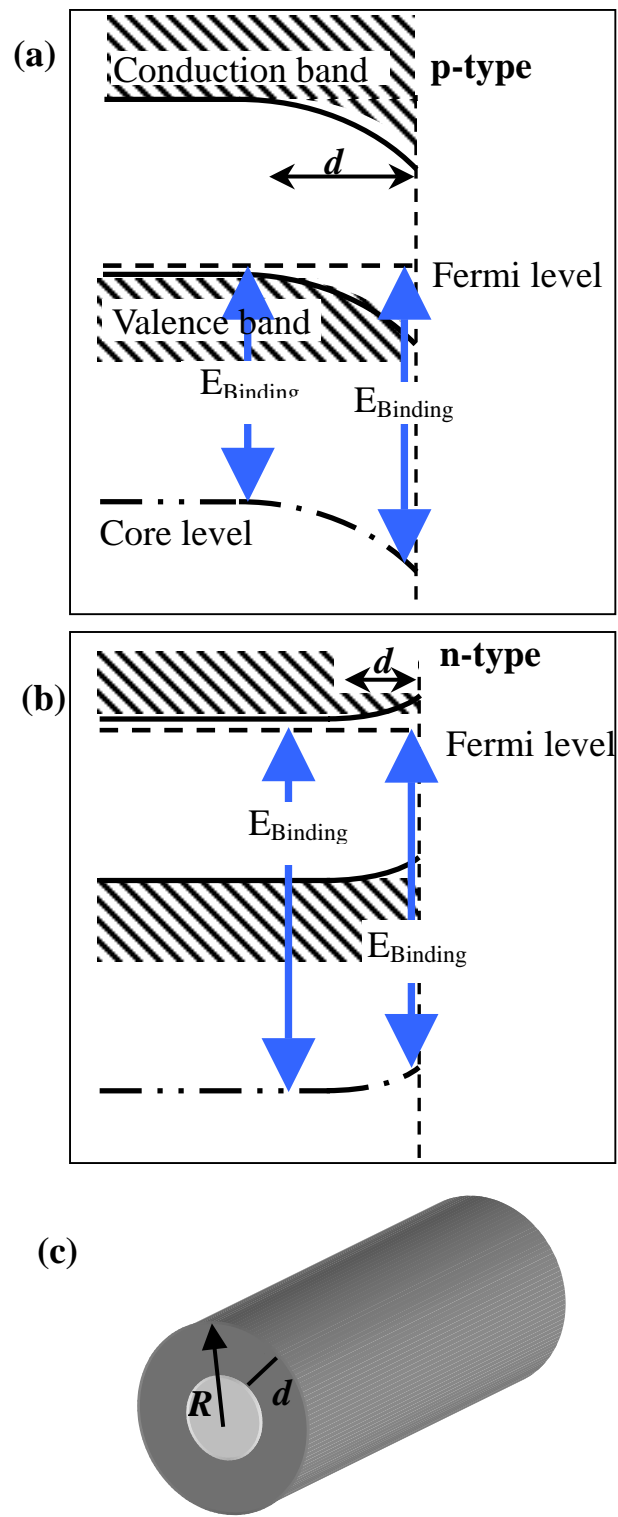

**(a)** p-type

Conduction band

$d$

Fermi level

Valence band

$E_{Binding}$    $E_{Binding}$

Core level

**(b)** n-type

$d$

Fermi level

$E_{Binding}$

$E_{Binding}$

**(c)**

$R$    $d$

**Figure 8**

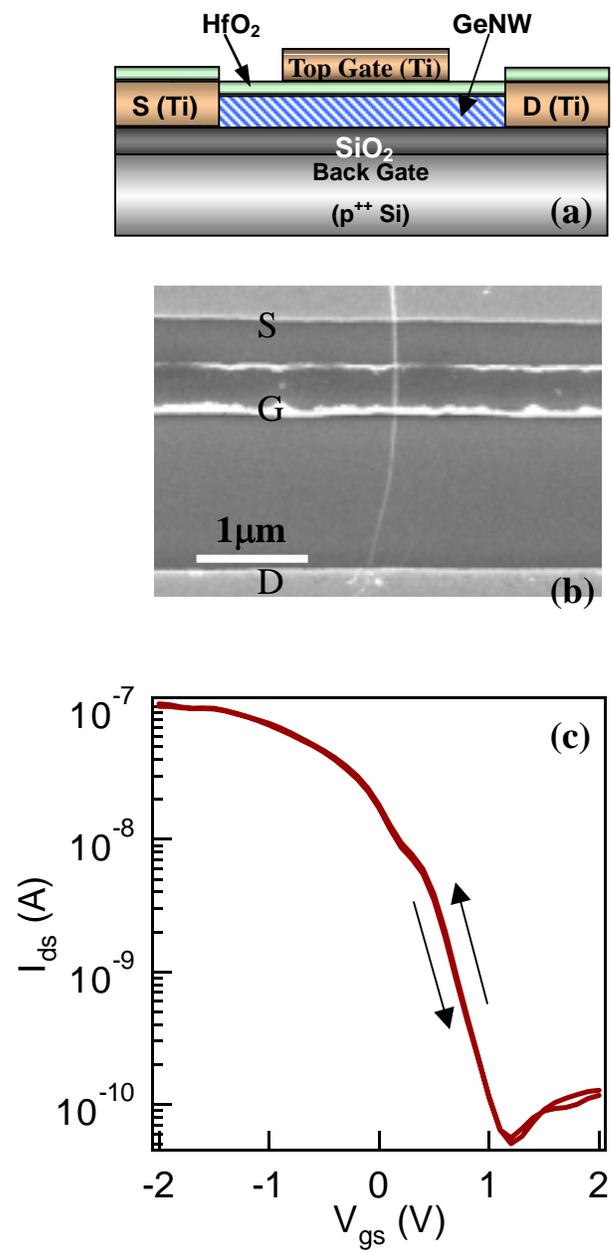

HfO$_2$

GeNW

Top Gate (Ti)

S (Ti)

D (Ti)

SiO$_2$

Back Gate

(p$^{++}$ Si)

**(a)**

S

G

1μm

D

**(b)**

$I_{ds}$ (A)

$V_{gs}$ (V)

**(c)**

**Figure 9**